\documentclass[fleqn,10pt]{wlscirep}

\usepackage{float}
\usepackage{framed}
\usepackage{lastpage}
\usepackage{bm}
\usepackage{physics}
\usepackage{amsfonts}
\usepackage[utf8]{inputenc}
\usepackage[T1]{fontenc}
\usepackage{mathpazo} 

\usepackage{amsmath}
\usepackage{soul}
\usepackage{xcolor}
\usepackage[symbol*]{footmisc}
\usepackage{todonotes}
\usepackage{graphicx}
\usepackage{booktabs}

\usepackage[utf8]{inputenc}
\newcommand{\rev}[1]{\textcolor{black}{#1}}

\DeclareRobustCommand{\myAA}{{\fontencoding{T1}\selectfont\char197}}

\title{\rev{Metrics for spin-based computing}}
 
\author[1,2,3,4,*]{Hidekazu Kurebayashi}
\author[5,*]{Giovanni Finocchio}
\author[6]{Karin Everschor-Sitte}
\author[7]{Jack C. Gartside}
\author[8]{Tomohiro Taniguchi}
\author[9]{Artem Litvinenko}
\author[4,9,10]{Akash Kumar}
\author[4,9,10,*]{Johan \myAA kerman}
\author[11]{Eleni Vasilaki}
\author[12]{Kemal Selçuk}
\author[12]{Kerem Y. Çamsarı}
\author[13]{Advait Madhavan} 
\author[3,4,10,*]{Shunsuke Fukami}

\affil[1]{London Centre for Nanotechnology, University College London,  17-19 Gordon Street, London, WC1H 0AH, United Kingdom}
\affil[2]{Department of Electronic and Electrical Engineering, University College London, Roberts Building, London, WC1E 7JE, United Kingdom}
\affil[3]{WPI Advanced Institute for Materials Research, Tohoku University, 2-1-1, Katahira, Sendai 980-8577, Japan}
\affil[4]{Center for Science and Innovation in Spintronics (CSIS), Tohoku University, 2-1-1 Katahira, Aoba-ku, Sendai 980-8577 Japan}
\affil[5]{Department of Mathematical and Computer Sciences, Physical Sciences and Earth Sciences, University of Messina, Messina, Italy}
\affil[6]{Faculty of Physics and Center for Nanointegration Duisburg-Essen (CENIDE), University of Duisburg-Essen, 47057, Duisburg, Germany}
\affil[7]{London Centre for Nanotechnology, Imperial College London, London, SW7 2AZ, United Kingdom}
\affil[8]{National Institute of Advanced Industrial Science and Technology (AIST), Research Center for Emerging Computing Technologies, Tsukuba, Ibaraki, 305-8568, Japan}
\affil[9]{Department of Physics, University of Gothenburg, Gothenburg 41296, Sweden}
\affil[10]{Research Institute of Electrical Communication (RIEC), Tohoku University, 2-1-1 Katahira, Aoba-ku, Sendai 980-8577 Japan}
\affil[11]{University of Sheffield, Sheffield, S10 2TN, United Kingdom}
\affil[12]{Department of Electrical and Computer Engineering, University of California, Santa Barbara, Santa Barbara, CA, 93106, USA}
\affil[13]{National Institute of Standards and Technology, Gaithersburg, MD, USA}
\affil[*]{Emails: h.kurebayashi@ucl.ac.uk, giovanni.finocchio@unime.it, s-fukami@riec.tohoku.ac.jp and johan.akerman@physics.gu.se}

\begin{abstract}
Spin-based computing is emerging as a powerful approach for energy-efficient and high-performance solutions to future data processing hardware. Spintronic devices function by electrically manipulating the collective dynamics of the electron spin, that is inherently non-volatile, nonlinear and fast-operating, and can couple to other degrees of freedom such as photonic and phononic systems. This review explores key advances in integrating magnetic and spintronic elements into computational architectures, ranging from fundamental components like radio-frequency neurons/synapses and spintronic probabilistic-bits to broader frameworks such as reservoir computing and magnetic Ising machines. We discuss hardware-specific and task-dependent metrics to evaluate the computing performance of spin-based components and associate them with physical properties. Finally, we discuss challenges and future opportunities, highlighting the potential of spin-based computing in next-generation technologies.
\end{abstract}

\begin{document}
\flushbottom
\thispagestyle{empty}
\maketitle

\section{Introduction} 
The modern age of information has been driven by the development and evolution of digital computers that utilize transistors as switches to encode deterministic Boolean/binary logic operations. These operations provide a firm foundation and clean abstractions on which the higher levels of the computing stack rest, allowing designers to implement programmable architectures for general-purpose algorithms in a technology-agnostic way, while still reaping the benefits of technological progress at the device level~\cite{hennessy2011computer}. However, technological developments based on conventional transistor-based computing face inherent limitations stemming from the end of Moore's law~\cite{shalf2020future}, particularly concerning power consumption~\cite{horowitz20141,jouppi2021ten} and scaling challenges~\cite{Ryckaert_NatRevEE2024}. In addition, the need for unconventional approaches beyond Boolean logic is growing, with the rise of data-driven algorithmic techniques such as machine learning~\cite{han2016eie,chen2016eyeriss,shao2019simba,reuther2020survey}, graph computations~\cite{shun2013ligra,abadal2021computing,low2014graphlab}, and combinatorial optimization~\cite{mohseni2022ising,yamaoka201520k,tatsumura2021scaling,lo2023ising}, with the aim to address emerging demands in high-performance computing and neuromorphic architectures where conventional general-purpose computing techniques become difficult to meet the computational demands of the future. 

The quest for best performance is fundamentally a quest for energy and time efficiency which involves a re-imagining of the computing stack from the algorithmic perspective while addressing the sources of inefficiency in the architectures~\cite{muralidhar2022energy,jouppi2021ten}. One primary source stems from a mismatch between the degree of parallelism that algorithms demand and what the hardware provides~\cite{hill2021accelerator}. \rev{For example, machine learning and graph algorithms often allow large data batches or independent sub-problems to be processed simultaneously (data and task parallelism), though certain operations must still occur sequentially due to inherent dependencies. This parallelism can take the form of pipelined execution of algorithmic stages, concurrent processing across multiple GPUs or CPUs, or distributed computing frameworks in cloud and data center environments.} In order to implement massively parallel, data-centric algorithms~\cite{Zha_arXiv2023,Bhatt_SciRep2024} efficiently, it is essential that the hardware mirrors the parallel nature of the algorithmic data flow~\cite{dally2020domain}. This requires careful designing of the memory hierarchy and the computational units it supports. Most architectures are not well suited for in-memory computation, leading to the energetically costly movement of data across units~\cite{sebastian2020memory}.  
Another source of inefficiency arises from the mismatch between the mathematical primitives required by the algorithms and the restricted ones provided by the hardware. Algorithms that rely on large quantities of random numbers, or those involving responses in dynamical physical systems — such as the locking of oscillators, or temporal dynamics of physical systems — are expensive to emulate with digital implementations~\cite{wright2022deep}. Tasks requiring temporal recurrency pose significant challenges for digital systems, as modelling recurrent behaviour often demands computationally intensive, step-by-step processing, making real-time execution costly. In contrast, physical systems—particularly magnetic materials with intrinsic memory-sensitive dynamics—naturally evolve their state in a continuous manner. Emerging technologies harnessing these inherent dynamics therefore offer potential advantages in real-time efficiency, while remaining compatible with existing complementary metal-oxide semiconductor (CMOS) technologies.

Spin-based computing has this potential~\cite{grollier2020neuromorphic, finocchio2021promise}. Magnetism offers a diverse range of ground states and dynamics, showcasing their non-volatility, time non-locality, rich and inherently nonlinear spin-wave dynamics and coupling to other degrees of freedom in e.g. electronics, photonics and phononics, electromagnetically and/or via spin-orbit interaction. This inherent nature allows them to be implemented in both existing digital computations and a wide variety of unconventional computing frameworks. They are scalable, routinely fabricated into nanometer-sized devices ~\cite{watanabe2018shape,behera2024ultra} and can be designed into electrically driven circuits that are compatible with existing CMOS technologies. Magnetic tunnel junctions (MTJs) are one example that has been successfully integrated into the back-end-of-the-line of a conventional CMOS process and are commercially available. 

One of the central aims of this Technical Review is to present key advances of spin-based computing schemes, with a specific focus on major developments in incorporating magnetic and spintronic elements into neuromorphic architectures. Our discussions further expand towards benchmarking of different approaches, starting from hardware-specific figures, such as energy consumption, speed, and device footprint, to architecture (task)-specific metrics to make fair comparisons across different neuromorphic computing systems. We conclude our article by offering a perspective view of key challenges and exciting opportunities ahead in particular research domains.

\section{Recent development and their working principle} 
\begin{figure}[t]
    \centering
    \includegraphics[width=1.0\linewidth]{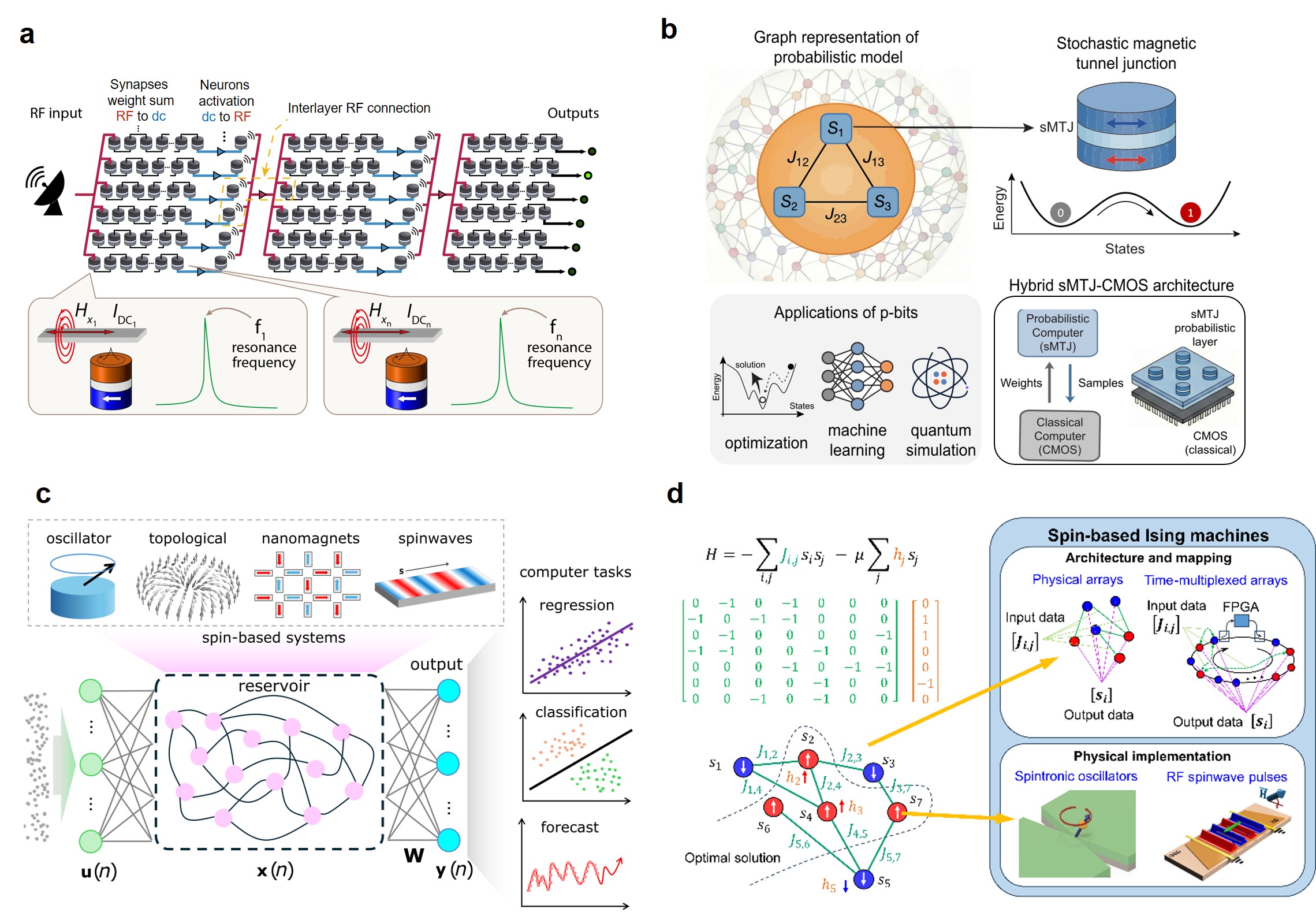}
  \caption{Four spin-based computing technologies discussed in this article. Conceptual diagrams and proof-of-concepts for \textbf{a} RF spintronic neural network, \textbf{b} probabilistic computer, \textbf{c} spin-based reservoir computer, and \textbf{d} spin-based Ising machine. Details of each system are discussed in the main text.}
  \label{fig1}
\end{figure}

\subsection{Radio frequency spintronic synapses and neurons}
Deep neural networks, which the recent progress in artificial intelligence 
relies on, are composed of non-linear activation functions (neurons) and trainable weighted sums (synapses) connected into a hierarchical layered structure. The major bottleneck in time and energy consumption when running and training neural networks on conventional digital CMOS-based hardware stems from the separation between memory and processing~\cite{Jangra_MMMED2024,Ryckaert_NatRevEE2024}.

Leveraging the non-volatile properties of spintronics to implement the synapses of neural networks is a promising path, exploiting a variety of spintronic technologies~\cite{grollier2020neuromorphic, finocchio2021promise} including spin-transfer torque magnetic random access memory (STT-MRAM), domain walls racetracks, skyrmionics, antiferromagnets and spin-Hall devices. \rev{Jung \textit{et al.}~\cite{Jung_Nature2022} and Borders \textit{et al.}~\cite{borders2024measurement} have demonstrated inference on crossbar arrays of monolithically integrated MTJ-CMOS cells with 4,096 and 20,000 MTJs respectively, where the former study achieved the power efficiency of 405 tera-operations per second per watt (TOPS~W$^{-1}$), competitive to other neural network accelerators~\cite{Guo_AcceleratorComparison}.} If the non-activation function between layers is implemented by digital circuits, there must be an analog-to-digital conversion at each layer, which becomes the next bottleneck in terms of energy and footprint. 
Consequently, there have been many proposals to use spintronic devices as analog neurons, again using a wide variety of devices~\cite{Roy_nrevelec2024,Rodrigues_PRAppli2023}. One of the major challenges is to interconnect the neurons and synapses at high density, for which most studies~\cite{Chakraborty_IEEE2020} have focused on crossbar arrays of synapses to vary the synaptic weights through the electrical conductance of the devices. 

Here, we highlight a solution which leverages the high-speed dynamics of spintronic devices. Ross \textit{et al.}~\cite{Ross_nnano2023} have demonstrated a deep neural network (Fig.~\ref{fig1}a) that emulate both synapses and neurons through successive radio-frequency(RF)-to-dc and dc-to-RF conversions using spintronic diodes and oscillators as illustrated in Fig.~\ref{fig1}a. A series of spintronic diodes ~\cite{Tulapurkar_Nature2005}, receiving RF inputs, plays the role of synapses by generating an output dc voltage, proportional to the input RF power and frequencies, which is the input of the oscillator acting as a neuron and performing the dc-to-RF conversion. The connection of those blocks as shown in Fig.~\ref{fig1}(a) represents the topology of the neural network.  The frequency selectivity of the spin diode effect makes it possible to chain synapses responding to different frequencies in series and to send the RF inputs at different frequencies in the chains. Compared to crossbar arrays, this frequency multiplexing makes the circuit architecture simpler (fewer connections) and limits sneak paths by physically separating each weighted sum (i.e.\ synaptic chain). 
Furthermore, this architecture is more flexible because the network topology relies on frequency matching between neurons and synapses rather than spatial arrangement, and accommodates different network topologies, e.g. fully connected, convolutional~\cite{Leroux_IOP2022} and sparse layers~\cite{Cai_APL2019}. The weights are controlled by the resonance frequency of the synapses, which can be tuned in a non-volatile way, both with  by using magnetic field from current lines (Fig.~\ref{fig1}a), and binary values by changing the vortex polarity or chirality.

Ross \textit{et al.}~\cite{Ross_nnano2023} achieved over 99\% accuracy on a drone classification task with real-life  RF signals without analog-to-digital converters. Removing the need for digitization provides massive gains in computing time and energy consumption for RF processing in edge devices, from radar and telecoms to biomedical applications. They estimate that 20 nm MTJs would consume 10 femtojoules per synaptic operation and 100 femtojoules per neural operation~\cite{Ross_nnano2023}. This estimation is based on the assumption that the energy loss in the resistive MTJs is dominant and that CMOS RF and dc amplifiers are needed between each layer of neurons and synapses to maintain signal levels throughout the network (see also Sec. 3.1). These estimations are comparable to or lower than those for memristive or photonic components, and several orders of magnitude less than current implementations of software neural networks~\cite{Chakraborty_IEEE2020}. In addition, RF synapses can be used for the computation of convolution~\cite{Leroux_IOP2022} and RF neurons can be coupled with memories for binary neural networks~\cite{Böhnert_CommEng2023}. 

Training is a critical challenge for all of these networks. For inference-only applications, the correct weight values need to be set in the hardware network. Training in software on conventional platforms, such as Graphics processing Units (GPUs), poses a challenge due to inherent discrepancies between the software model and the actual hardware network implementation (e.g., spintronic or other specialized networks). This requires methods of noise-aware training~\cite{Kariyappa_IEEE2021} or chip-in-the-loop training~\cite{Buckley_nanophoto2023}. For on-chip learning applications, the issue is to develop training algorithms that are adapted to a hardware network, including the non-idealities of its components, while conserving state-of-the-art accuracy. The growing interest in artificial intelligence dedicated hardware, as well as the desire to find bio-plausible learning algorithms, has led to many proposals of hardware-friendly algorithms often based on physical principles~\cite{Changze_arX2024}. Co-designing such algorithms together with spintronics neural networks is a challenging but exciting goal.

\subsection{Spintronic p-bits}

Stochasticity in physical systems, although problematic for deterministic algorithms, can be a powerful computational resource for applications such as Monte Carlo methods, optimization, and artificial intelligence. Although many physical systems can provide noise as computational entropy, the physics of superparamagnetism combined with the read-out capabilities of stochastic MTJs, or s-MTJs, may be particularly advantageous for high-speed throughput in true random numbers generation. This functionality has been used to experimentally demonstrate the concept of probabilistic-bit (p-bit)~\cite{PhysRevX.7.031014,borders2019integer} (see Fig.~\ref{fig1}b), opening a path to the hardware implementation of Richard Feynman's suggestion in his famous lecture titled “Simulating Physics with Computers” where he stated that “\textit{… the other way to simulate a probabilistic nature … might still be to simulate the probabilistic nature by a computer which itself is probabilistic…}”~\cite{Feynman1982}. 

p-bits have proven effective for emulating the transverse field Ising model —a sign-problem--free system employed in D-Wave quantum annealers~\cite{camsari2019scalable}. However, attempts to generalize this approach to universal quantum circuits encounter the Monte Carlo sign problem, limiting straightforward sampling in non-stoquastic Hamiltonians~\cite{chowdhury2023emulating}. Consequently, while p-bits offer an efficient hardware platform for certain optimization and sampling tasks, they cannot universally replace qubits. Nevertheless, their application domain overlaps significantly with that of qubits, focusing on quantum-inspired algorithms rooted in optimization and sampling.

Furthermore, the s-MTJ-based p-bit can beneficially replace pseudo-random number generators in current CMOS-based hardware, given that the projected s-MTJ-based p-bit requires roughly four order of magnitudes less transistors than a high-quality CMOS  pseudo-random number generator ~\cite{singh2024cmos}.

Building on several simulation-based studies~\cite{PhysRevX.7.031014, sutton2017intrinsic}, proof-of-concept demonstrations of s-MTJ-based probabilistic computers have showcased functionalities including invertible logic~\cite{borders2019integer,lv2019experimental}, combinatorial optimization~\cite{borders2019integer,10019520,si2024energy}, machine learning~\cite{mizrahi2018neural,kaiser2022hardware, li2024restricted}, inference~\cite{singh2023hardware}, and quantum simulation~\cite{10019530}. For instance, Borders \textit{et al.}~\cite{borders2019integer} demonstrated integer factorization as an example of such optimization, using eight p-bits (neurons) connected by a microcontroller (synapses). More recently, Si \textit{et al.}\cite{si2024energy} scaled the system to 80 p-bits to address travelling salesman problems. Although these proof-of-concept systems have been constrained by limited p-bit counts and slow random number generation, approaches such as software-based simulations of s-MTJs \cite{PhysRevApplied.17.024052,onizawa2024enhanced}, digital circuit emulations~\cite{10185207,niazi2024training}, and heterogeneous s-MTJ/CMOS systems~\cite{singh2024cmos,singh2023hardware} have exhibited promising performance.

In parallel with these proof-of-concept systems, extensive research has focused on understanding the stochastic behaviour of nanomagnets and on developing s-MTJs with properties favourable for practical applications (Fig.~\ref{figMTJ}). Among the most critical metrics of s-MTJs is the speed of random magnetization switching, that is, the rate of random telegraph noise generation, which directly influences computation speed and time-to-solution. Experiments have observed random telegraph noise on the nanosecond timescale in in-plane easy axis s-MTJs~\cite{PhysRevLett.126.117202, safranski2021demonstration, PhysRevApplied.20.024002}. Beyond switching speed, other key considerations for probabilistic computers include  robustness to external fields, bias-voltage dependence and temperature sensitivity. To tackle these challenges, various s-MTJ design have been explored, including free layers with synthetic antiferromagnetic structures~\cite{PhysRevApplied.18.054085, PhysRevB.108.064418, PhysRevApplied.21.054002} and double free-layer configurations~\cite{PhysRevApplied.15.044049, PhysRevApplied.21.054002, 10.1063/5.0219606}. In addition, three-terminal MTJ structures controlled with spin-orbit torque have been developed as alternatives to the two-terminal MTJs commonly employed in current MRAM technologies~\cite{8695882, 10019520}. Furthermore, electrical coupling of multiple s-MTJs has been investigated for compact and energy-efficient implementations\cite{debashis2020correlated, talatchian2021mutual, schnitzspan2023electrical, gibeault2024programmable}.

Lastly, alternative options for constructing spintronic p-bits include thermally stable two- or three-terminal MTJs driven by  electric current~\cite{Fukushima_2014, shao2021implementation, PhysRevApplied.19.024035, ren2024initialization}, voltage-controlled magnetocrystalline anisotropy\cite{WOS:000912390600001}, or strain\cite{WOS:000669542400008}, which  can produce switching with arbitrary probabilities, as well as phase noise in spin-torque nano oscillators (STNOs)\cite{PhysRevApplied.21.034063}. 
In addition, numerous non-magnetic solid-state systems with inherent probabilistic behavior have been explored, such as resistive random access memory devices~\cite{mi13060924}, memristors\cite{woo2022probabilistic}, perovskite\cite{park2022efficient}, ferroelectric devices\cite{10149538}, avalanche and Zener diodes\cite{whitehead2023cmos,patel2024pass}, and microelectromechanical systems~\cite{wutrue}.

\subsection{Magnetic reservoir computing}

Modern machine learning approaches, particularly deep learning, have achieved remarkable success in answering complex tasks by relying on vastly expanded neural networks that require huge training costs, both time and energy. While they perform very well after being trained with large-scale parameter optimisation, they often lack the capability for real-time adaptation to temporally changing datasets due to both slow training speeds and susceptibility to catastrophic forgetting~\cite{French_TCS1999}. Smaller-scale, lower-complexity tasks with time-series data represent a significant and growing portion of artificial intelligence demand, e.g. in `edge computing' applications where real-time processing is required at a local device without data transmission to a cloud/remote data centre. Reservoir computing (RC)~\cite{Nakajima_RCBook2021}, a computational scheme that initially emerged as a variant of recurrent neural networks, is a promising candidate for addressing these needs. Stemming from echo state networks introduced by Jaeger \textit{et al.}\cite{Jaeger_Rep2001} and liquid state machines by Maass \textit{et al.} ~\cite{Maass_NeurCom2002} independently, the field of RC has advanced considerably with numerous proof-of-concept demonstrations for both software-based~\cite{Sun_IEEE2024} and physical-based~\cite{TANAKA_NeurNet2019,10.1063/5.0148469, Liang_NComm2024, allwood2023perspective, Yan_NComm2024}. 

RC features a distinctive architecture, where the central component—the reservoir—is a complex, nonlinear system characterised by fading memory (see Fig.~\ref{fig1}c). The reservoir implements a nonlinear transformation of input data, projecting them into a high-dimensional state space. A simple, typically linear regression layer then decomposes this high-dimensional representation, generating the desired computational response. This structure significantly reduces the number of connection weights requiring optimisation as only those to the output layer are trained, thereby enabling fast adaptation to spatiotemporal data. \rev{Cisneros et al. found that reservoir computing models can achieve remarkable learning efficiency and, on their benchmarking, outperform more standard supervised models including a gated RNN, i.e. long short-term memory (LSTM), and Transformers~\cite{Cisneros_ConfLLA2022}. Manneschi et al. compared a variant of a reservoir model with an LSTM under parameter-matched conditions on the permuted sequential MNIST task and achieved superior performance~\cite{Manneschi_IEEETransNNLS2023}.} To unlock RC’s full potential and appeal to the industry, further research is needed to develop RC systems tailored to specific applications.

A reservoir must exhibit nonlinearity and fading memory, with sufficient complexity to capture rich dynamics. For a physical reservoir, the nonlinear response to input signals, combined with high-dimensional mapping between input signals and reservoir nodes, is essential for transforming complex nonlinear problems into simpler solvable computations in the final output layer. The fading memory property is in particular relevant for time-series tasks. In such cases, the training of conventional neural networks becomes more challenging and computationally expensive - requiring techniques such as backpropagation through time. Reservoir approaches have been shown to offer considerable benefits, simplifying training and allowing efficient methods such as regression or gradient-based techniques~\cite{manneschi2021sparce} to solve impressively challenging tasks.

Fading memory, which is crucial for time-dependent tasks, can be mathematically expressed by defining the state vector $x(n)$ at the $n$-th time step is defined by a series of the input signal $u(i)$ as:
\begin{equation}
    x(n) = f(u(k), u(k+1), u(k+2), \ldots, u(n-2), u(n-1), u(n)).
\end{equation}
Note that due to the fading memory, $x(n)$ is only a function of limited inputs between the $k$-th and $n$-th time steps. This property provides computational capacity for tasks with time-dependent data, such as future prediction. The timescale of the fading memory can be optimised to suit different tasks, and hierarchical structures can tackle complex problems or improve performance~\cite{manneschi2021exploiting}.

The first realisation of spin-based physical reservoir computers was reported by Torrejon \textit{et al.}~\cite{Torrejon_Nature2017} where nanoscale spintronic oscillators enabled spoken-digit recognition in an RC scheme.
In this work, the system's complexity needed to be artificially enhanced through time multiplexing, however, since then, there have been several spin-based RC demonstrations~\cite{10.1063/5.0148469}, e.g.~using STNOs~\cite{Kanao_PRAppl2019,Tsunegi_JJAP2018,Akashi_PRRes2020,Wu_PRMater2023,Tsunegi_AdvIntSys2023}, magnetic domain walls~\cite{Ababei_SciRep2021,Zhou_SciAdv2025}, spin-waves~\cite{Watt_PRAppl2021,Nakane_PRApp2021,Namiki_AdvIntSys2023,Namiki_NeurCompEng2024,Iihama_npjspin2024,Nagase_PRAppl2024}, magnetic nanoarrays including artificial spin ices~\cite{Gartside_NNano2022, stenning2024neuromorphic, hu2023distinguishing,hon2021numerical, jensen2018computation} and nanorings~\cite{vidamour2023reconfigurable, dawidek2021dynamically}, magnetic vortices~\cite{Körber_NComm2023} and magnetic skyrmions~\cite{PhysRevApplied.9.014034, Pinna_PRApp2020, Raab_NComm2022, Yokouchi_SciAdv2022, Lee_NMater2024, Lee_SciRep2023, Msiska_AdvIntSys2023, everschor2024topological}. Recent review articles provide excellent summaries of RC implementations in physical  systems~\cite{TANAKA_NeurNet2019,10.1063/5.0148469,Cucchi_IOP2022, Liang_NComm2024, allwood2023perspective, Yan_NComm2024}, together with a python package for processing of physical reservoir computing~\cite{youel_2024arXiv}. 

Magnetic systems offer several attractive benefits for physical RC. The magnetic free energy is determined by a multitude of magnetic anisotropies and applied magnetic field, providing magnetic hystereses and rich domain configurations. The elementary excitation of magnetic orders, \textit{i.e.} magnons, has a distinctive dispersion which can be controlled by the magnetic-dipole and exchange interactions, providing a computational resource due to its intrinsic nonlinearity~\cite{Gurevich_book2020}. As a result, when driving a magnetic system with a physical input, such as magnetic field and spin torques~\cite{RALPH2_JMMM2008,Mancon_RMP2019}, the output is generally both nonlinear and recurrent. These unique properties differentiate from other physical systems explored for neuromorphic computing, \textit{e.g.} photonics where physical memory and nonlinearity are more difficult to generate and engineer intrinsically. 
In addition, magnetic systems can support a wide range of metastable states, each exhibiting rich dynamics for computational responses. This enables the creation of more complex reservoirs with enhanced computational capacity. Additionally, a single physical system can be reconfigured to produce different reservoir responses, allowing it to be tailored to solve various tasks~\cite{Lee_NMater2024}. Not only can adjusting its microstate improve a reservoir's performance, but also combining different magnetic materials and incorporating multiphysics effects\cite{everschor2024topological} expands the amount of inputs a reservoir can process, further increasing its computational capacity.

\subsection{Magnetic Ising machine}

Traditional computational methods struggle with so-called Combinatorial Optimization problems (COPs), as the number of possible combinations grows factorially with the problem size. Common COPs include resource allocation~\cite{xie2013optimal}, scheduling~\cite{yang2011optimal}, graph partitioning~\cite{barahona1988application,ushijima2017graph}, drug design~\cite{blunt2022perspective, wu2018moleculenet}, routing in logistics~\cite{laporte1992vehicle, neukart2017traffic}, designing efficient networks~\cite{fernandez2018metaheuristics}, and even enhancing machine learning and artificial intelligence models~\cite{bohm2022noise,laydevant2024training}. To address these challenges, heuristic methods have been explored, where Ising Machines (IMs)\cite{Johnson2011nature, Honjo2021SciAdv100kCIM, bohm2019poor,wang2019oim,cen2022large} attract particular attention as a wide range of COPs can be mapped~\cite{lucas2014ising} onto the Ising model, with a Hamiltonian of the form,
\begin{equation}
    \label{eq:hamiltonian}
    H (s_1,...,s_N) = -\sum_{i<j} J_{ij} s_i s_j - \sum_{i=1}^N h_i s_i 
\end{equation}
where $s_i=\pm 1$ are the Ising spins, $N$ is the total number of spins, $J_{ij}$ is their coupling matrix, and $h_i$ is a local biasing field. 
An Ising machine is any physical system~\cite{mohseni2022ising} that minimizes the Ising Hamiltonian (Eq.2), where the solution to the corresponding COP is encoded in the spin states at the minimum. 
Utilizing intrinsic dynamics and various annealing schemes, IMs evolve over time to their lowest-energy binarized spin state, representing an optimal solution to the COP.
IMs can be implemented in a wide range of different hardware, broadly categorized into spatial networks on the one hand, such as commercial quantum annealer\cite{Johnson2011nature} and classical oscillator networks \cite{wang2019oim}, and time-multiplexed networks on the other, such as the optical Coherent IM\cite{Honjo2021SciAdv100kCIM}. Among various physical implementations of equivalent Ising spins, spintronic oscillators and spin wave-based propagating RF pulses stand out due to their potential for high speed, scalability, and energy efficiency~\cite{albertsson2021ultrafast,houshang2022phase,gonzalez2024spintronic,litvinenko2023spinwave}. 

Spatial IMs based on both STNOs~\cite{albertsson2021ultrafast} and spin Hall nano-oscillators (SHNOs)~\cite{houshang2022phase,mcgoldrick2022prappl,liu2024advancing} have been simulated and partially demonstrated. When subjected to second harmonic injection locking (SHIL) at twice their intrinsic frequency, the individual phases of the oscillators in the network become bistable, taking on only binarized values of either 0 or $\pi$ rad, as depicted in Fig. 1d, thereby realizing the artificial spin states ($s_i=\pm 1$) for mapping the Ising Hamiltonian. SHNOs can be made as small as 10 nm and made to mutually synchronize in networks as large as 100,000 nano-oscillators~\cite{behera2024ultra,behera2025ultra}, and the inter-SHNO coupling strength $J_{ij}$, and sign, can be programmed using memristive voltage gates~\cite{zahedinejad2022memristive,kumar2025spin}. Their very high operating frequencies, direct physical interactions through exchange, spin waves, and dipolar coupling, and their ultrafast relaxation rates, offer the potential for ultrashort time-to-solutions, making them a promising candidate for further development.

Another approach is time-multiplexed spin-wave IMs (SWIM)~\cite{litvinenko2023spinwave}. In SWIM, each Ising spin is represented by spin-wave RF pulses, which undergo phase binarization through an external microwave phase-sensitive amplifier before propagating along a YIG delay line. Losses in the delay line are compensated through linear amplification in a loop circuit, allowing continuous circulation of RF spin-wave pulses. The phase binarized RF pulses act as artificial Ising spin states, $s_i=\pm 1$, with coupling controlled either physicially—by delaying a portion of each RF pulse in additional delay lines—or digitally, using Field Programmable Gate Array (FPGA)-based measurement and feedback blocks. Thanks to the low propagation speed of spin waves compared to their optical counterparts~\cite{Honjo2021SciAdv100kCIM}, SWIMs can be ultra-small, exhibit high thermal stability, and offer excellent scalability.

Notably, in physical oscillator array IMs, the relaxation does not require any computational control while in the time-multiplexed case, it only requires a simple and fast matrix multiplication operation. This is a core advantage over von Neumann architectures and allows for both acceleration of the computational speed and a corresponding reduction in power consumption~\cite{gonzalez2024spintronic}. Recent advancements in nanofabrication and spintronic materials have enabled the development of highly integrated SHNO arrays ~\cite{zahedinejad2020two,behera2025ultra}, further control of coupling with spin-waves and non-volatile gating put forward these systems for large scale, truly nano-scopic IMs.  
At the same time, enhancements in coherence, stability, and nanoscale spin-wave waveguides~\cite{divinskiy2021dispersionless} support the advancement of time-multiplexed propagating SWIMs. These achievements are paving the way for practical implementation of large-scale spintronics-based IMs capable of solving COPs that require real-time rapid solving such as high frequency trading~\cite{tatsumura2023real} and vehicle self-driving.

\section{Metrics}

\subsection{Radio frequency spintronic synapses and neurons}

The research in neuromorphic spintronics is emerging thanks to the built-in functionalities of spin-based devices making them very promising for the compact and efficient hardware implementation of synapses and neurons ~\cite{grollier2020neuromorphic, finocchio2021promise}. As already pointed out, among the possible directions, the implementation of neuromorphic spintronic architectures based on RF synapses and neurons is emerging as a promising alternative to the crossbar geometries enhancing the advantages of spintronics in terms of energy efficiency, scalability, CMOS compatibility and velocity with the intrinsic RF characteristics of MTJs able to generate, process and detect RF signals\cite{Ross_nnano2023,Zeng_PEAppl2024}. Below we introduce an overview of potential metrics and performance of the MTJs for evaluating RF neuromorphic spintronics.

\textit{Synapses. } The synaptic operation is based on the RF to dc conversion driven by a network of MTJs that are connected in series and provide the diode rectification effect~\cite{Finocchio_APL2021}. The main metric here is the energy per synaptic operation. The work proposed by Ross \textit{et al.}~\cite{Ross_nnano2023} considered spin diodes working in a passive regime where only RF current is applied directly to the MTJ. Depending on the synaptic weight tunable by the field generated by a DC current flowing in a current line positioned on top of the MTJ, the rectified voltage can be of the order of a few µV for input microwave powers at µW level. The total power dissipation is then given by two contributions, the one directly dissipated in the spin-diode chains because of the RF current and the one necessary to tune the synaptic weights by the external DC current used to generate the field applied to the device. The final  energy per synaptic operation will depend on the transient speed of the spin-diodes to reach its stationary response which is 5 to 6 times the inverse of the resonance frequency as estimated by micromagnetic simulations~\cite{Mazza_PRAppli2022}.

There are several challenges to face here before this approach can be competitive with state-of-the-art solutions. First of all, passive spin-diodes have rectification curves exhibiting positive and negative voltages at different frequencies hence small variations of the parameters (i.e. the resonance frequency) can reduce the energetic performance of the synaptic operation. Another important improvement to reduce the energy per synaptic operation will be the enhancement of the spin diode sensitivity by applying together to the RF signal a DC current and using spin diodes working in an active regime. For example, by using active spin-diodes with the injection locking mechanism the sensitivity can reach a value of 4 MV/W~\cite{Jiang_APRev2024,Goto_nnano2019} more than 3 to 4 orders of magnitude larger than passive configurations~\cite{Fang_ncomm2016,Finocchio_APL2021}. At the device level, the main characteristics to improve the scalability and increase the maximum number of synapses per neuron are the selectivity of the spin-diodes and the Resistance-Area of the MTJs allowing to have a larger number of devices connected in series. The former property can be improved by reducing the Gilbert damping or working with active spin-diodes while the latter needs a large effort from a deposition process development point of view. A three-terminal MTJ can be used to remove the need of the current line and then having a reduction of area occupancy and energy consumption. In this latter configuration,  the dc spin-orbit torque or the voltage-controlled magnetocrystalline anisotropy can be used for tuning the synaptic weight, i.e. the resonance frequency of the spin diodes~\cite{Finocchio_APL2021}.

\textit{Neurons.} The rectified output of each synaptic chain is then applied to the RF neuron, an MTJ working as a STNO, which converts a dc input to an RF output. The important metric here is the power consumption per neuron which evaluates the total power dissipated by each RF neuron in the network during its operation. One important aspect is that in order to generate an RF signal the $I_\text{DC}$ biasing the STNO should be larger than a critical value. This local bias is the main contribution to the power consumption per neuron, that is given by the 0.5 <$R_\text{MTJ}$>$I_\text{DC}^2$ where <$R_\text{MTJ}$> and $I_\text{DC}$ are the average resistance of the MTJ and the current flowing on it.   In the proposal by Ross \textit{et al.}~\cite{Ross_nnano2023}, the RF output of the neuron has to be amplified to feed to the next layer of the RF neural network. This need can be mitigated by developing STNOs with larger microwave emissions or connect them in a network~\cite{Jiang_APRev2024} or with spintronic amplifiers\cite{Goto_nnano2019,Zhu_ncomm2023}.

At the network level, RF neuromorphic spintronics can offer a reduced latency per inference~\cite{behera2024exploring} that is set by the resonance frequency of the synapses and the oscillation frequency of the neurons. From a perspective point of view considering a working frequency of 5 GHz with 1 ns transient for the synapses and neurons monolithically integrated with CMOS, the amplifier and the signal propagation delays, the time to complete a single inference task can be around a few tens of ns for each network layer. Of course, the time to complete a single inference task depends on the number of layers; however, this direction is promising as compared to ms latency per inference typical in state-of-the-art neural networks~\cite{behera2024exploring}. Combining these ideas with experimental validations and monolithic co-integration with CMOS will pave the way for concrete advancements in RF neuromorphic spintronics, enabling their future deployment as neuromorphic accelerated hardware.

\subsection{Spintronic p-bits}

\begin{figure}[t!]
    \centering
    \includegraphics[width=1.0\linewidth]{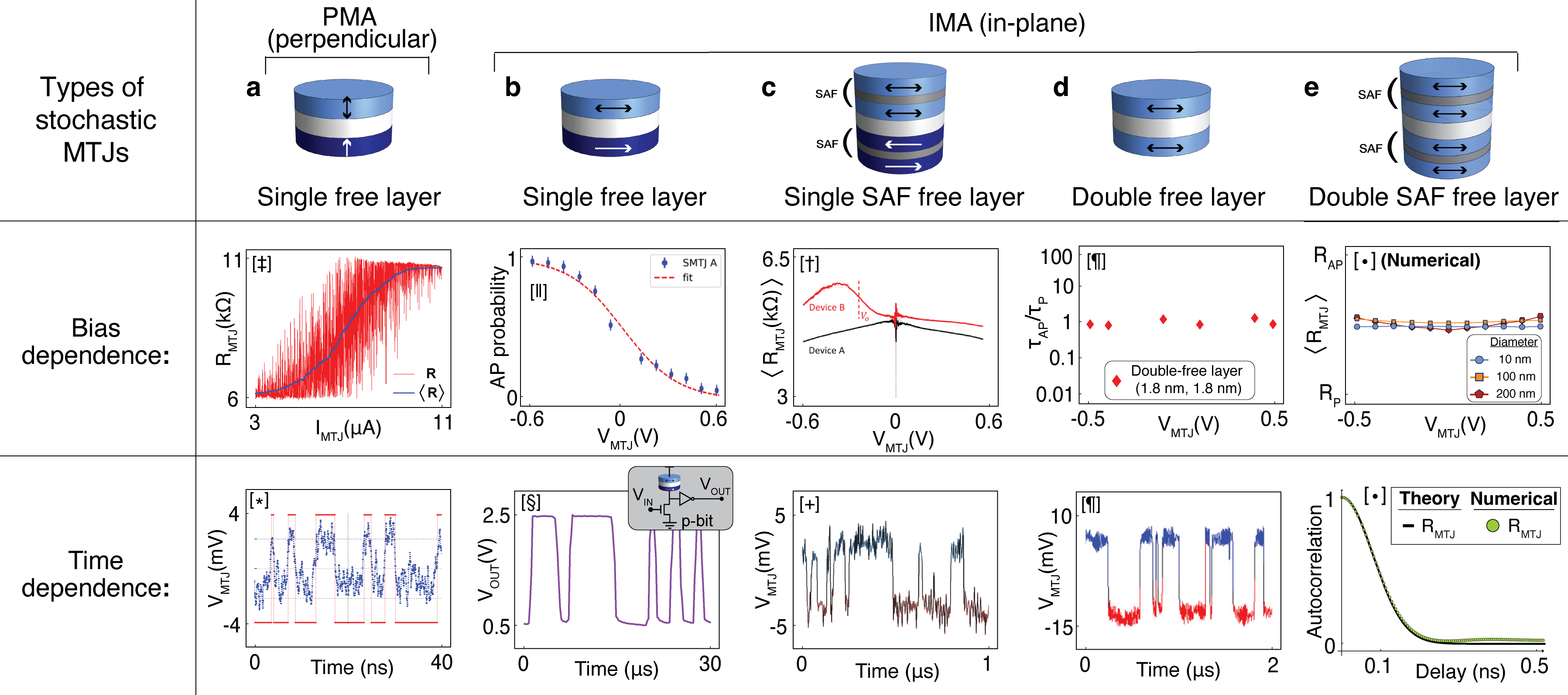}
  \caption{Types of stochastic magnetic tunnel junctions (MTJ) designs. \textbf{a} A fixed‐free MTJ with perpendicular magnetic anisotropy (PMA) exhibits bias dependence $[\ddagger]$\cite{10019530} and can reach nanosecond‐scale fluctuation speeds $[\ast]$\cite{soumah2024nanosecond}. \textbf{b} A fixed‐free MTJ with in‐plane magnetic anisotropy also shows bias dependence $[\Vert]$\cite{schnitzspan2023electrical}. This type of sMTJ is combined with CMOS to build a p-bit demonstrating 50/50 probability fluctuations on the microsecond scale [§]\cite{singh2023hardware}. \textbf{c} MTJs incorporating synthetic antiferromagnets (SAF) in both the fixed and free layers can be either bias‐dependent or bias‐independent, depending on the extent of in‐plane magnetic anisotropy$[\dagger]$\cite{sun2023stochastic}.  Here, $\langle R_{\mathrm{MTJ}} \rangle$ denotes the DC resistance 
$R_{\mathrm{dc}} = V_{\mathrm{MTJ}}/I_{\mathrm{MTJ}}$, measured with a bandwidth much lower than the fluctuation frequency, so that fast resistance fluctuations are time-averaged. Alternatively, pairing a non-SAF fixed layer with an SAF free layer offers robustness against external fields and sub‐microsecond switching $[+]$\cite{PhysRevApplied.18.054085}. \textbf{d} Removing the fixed layer yields a double‐free MTJ, which is effectively bias‐independent, as evidenced by the fraction of time spent in antiparallel (AP) vs. parallel (P) states under any bias, with sub‐microsecond dynamics $[\P]$\cite{10.1063/5.0219606}. \textbf{e} Theoretically, an SAF design for both free layers is proposed to be bias‐independent and free of dipolar coupling for any device diameter, exhibiting nanosecond time scales $[\bullet]$\cite{PhysRevApplied.21.054002}.}
  \label{figMTJ}
\end{figure}

Computing architectures using spintronic p-bits can be categorized into analog and digital approaches. The optimal device metrics of s-MTJs strongly depend on the chosen architecture and should be determined through co-design considerations. In the analog approach, p-bits are connected via a synaptic-weight circuit, which can be implemented with CMOS or emerging technologies. This setup represents an asynchronous, free-running architecture~\cite{camsari2023fullstack}. In the digital approach, s-MTJs serve as high-speed sources of random bits, driving large-scale digital circuits. By leveraging the fine-grained reconfigurability of FPGAs, researchers have demonstrated energy-efficient probabilistic computing architectures, incorporating sparsity engineering to optimize performance~\cite{singh2024cmos}. In the following, several critical device metrics are discussed.

\subsubsection*{Random Number Generation Speed and Autocorrelation}
The fluctuation timescale of magnetization direction in s-MTJs determines the speed of random bit generation, a primary metric for probabilistic computing (Fig.~\ref{figMTJ}). Faster random switching (or the rate of random telegraph noise generation) leads to shorter computation times. However, the response time of synaptic-weight circuits and delays in peripheral circuits must also be considered. Recent experimental studies have achieved fluctuation speeds in the range of microseconds to nanoseconds using in-plane easy-axis MTJs (Fig.~\ref{figMTJ}b), corresponding to bit generation rates of MHz to GHz~\cite{PhysRevLett.126.117202, safranski2021demonstration, PhysRevApplied.20.024002}. A p-bit circuit using in-plane MTJs produces output fluctuations of microsecond timescale as well (bottom row in Fig.~\ref{figMTJ}b). Understanding the physics governing the relaxation time of the magnetization direction is essential to optimize this speed~\cite{Kanai_PRB2021}. Another measure of speed can essentially be how fast random fluctuations lose their memory by evaluating the autocorrelation (bottom row in Fig.~\ref{figMTJ}e).

\subsubsection*{Power Consumption and Scaling Considerations}

Recent studies estimate that each p-bit consumes approximately 20 µW when implemented with fast s-MTJs~\cite{Hassan_PRAppl2021}. At this power level, a system with one million p-bits would require around 20 W for p-bit operations alone. In parallel,  similar power numbers were estimated for specialized analog synapses \cite{sutton2020autonomous}, suggesting that a fully integrated probabilistic computing system at this scale would operate below 100 W~\cite{sutton2020autonomous}. 

While these power levels are feasible for ambitious implementations with up to million bits in single chips, power consumption is expected to be a primary limitation for scaling beyond this regime. Achieving larger-scale probabilistic systems will require either reducing the power per p-bit or implementing architectural techniques, such as selectively powering down inactive parts of the system to manage overall energy consumption~\cite{camsari2023fullstack}. At the device level, it is of significant to achieve the aforementioned fast fluctuation at smaller bias current and voltage.

\subsubsection*{Noise Amplitude, Resistance Nature and Bias-Voltage Dependence}
A key advantage of s-MTJs is their large signal amplitude during switching: resistance fluctuations produce 100 mV –200 mV voltage swings, significantly higher than typical thermal noise in CMOS circuits~\cite{camsari2023fullstack} at room temperature. Thanks to the collective dynamics of magnetization switching combined with high tunneling magnetoresistance, the resulting voltage fluctuations are naturally large enough to drive surrounding CMOS circuits, eliminating the need for the extra amplification circuitry required in other probabilistic circuit designs~\cite{cheemalavagu2005probabilistic,patel2024pass}.

 Beyond speed and power, the nature of resistance fluctuations is crucial for p-bit operation. The switching dynamics of s-MTJs can exhibit continuous or telegraphic noise, with a behaviour that is dependent on the bias voltage~\cite{Hassan_PRAppl2021}. The spin-transfer-torque effect on the fluctuations can be observed as a sigmoidal response in time average (middle row in Fig.~\ref{figMTJ}a,b). Considering the minimal CMOS designs for the p-bits, bias-voltage independence is desirable ~\cite{Camsari_IEDL2017, Hassan_PRAppl2021, singh2024cmos}, which can be achieved by minimizing the spin-transfer torque effect in s-MTJs. One technique is the use of double-free-layer MTJs~\cite{PhysRevApplied.15.044049,10.1063/5.0219606, PhysRevApplied.21.054002}, which enables bias-free devices while preserving the desired stochastic behaviour (Fig.~\ref{figMTJ}d,e).

\subsubsection*{Robustness to Magnetic Field and Temperature}
For practical probabilistic computing, ensuring magnetic immunity is critical, as external magnetic fields can disrupt device operation~\cite{Dieny_IEEE2024}. Various nanomagnet design strategies have been explored to enhance robustness. One promising approach involves employing synthetic antiferromagnetic structures in the free layer, which significantly improve resilience against external perturbations~\cite{PhysRevApplied.18.054085, sun2023stochastic, PhysRevB.108.064418}. Another issue is the reliability emerging from the internal magnetic fields where closely placed nanomagnets get stuck because of their dipolar interactions. A technique addressing this issue is the use of double-free-layer MTJs with synthetic antiferromagnetic structures, which effectively reduces the dipolar coupling at any device diameter while preserving the desired stochastic behaviour (Fig.~\ref{figMTJ}e)~\cite{PhysRevApplied.21.054002}. In addition to the magnetic field, insensitivity of properties of stochastic MTJs against the temperature variation is also an important metric and understanding the mechanism governing the temperature dependence~\cite{kaneko2024temperature, Elyasi_PRB2024} should be of significance for broadening the range of applications.

\subsection{Magnetic reservoir computing}

Magnetic reservoir computing metrics can be categorized into task-independent measures, which assess the intrinsic computational properties of the reservoir, and task-dependent measures, which evaluate its effectiveness in solving specific computational problems. \hyperref[LastPage]{Supplementary Table 1} summarises these metrics evaluated in several studies on magnetic reservoir computing \rev{and further metrics are listed as Supplementary Note 1}.

\subsubsection*{Task-independent metrics}

Task-independent metrics\cite{jaeger2002tutorial,love2023spatial,Dambre_SciRep2012} evaluate the intrinsic properties of the reservoir, focusing on fundamental characteristics such as non-linearity, memory capacity, and dimensionality expansion, independent of specific computational tasks. These metrics are essential for characterising the computational potential of physical reservoirs and optimising their performance across diverse applications.
Importantly, non-linearity and memory capacity can be evaluated spatially resolved providing insights into the local computational properties of the reservoir.

\begin{figure}[t]
    \centering
    \includegraphics[width=1.0\linewidth]{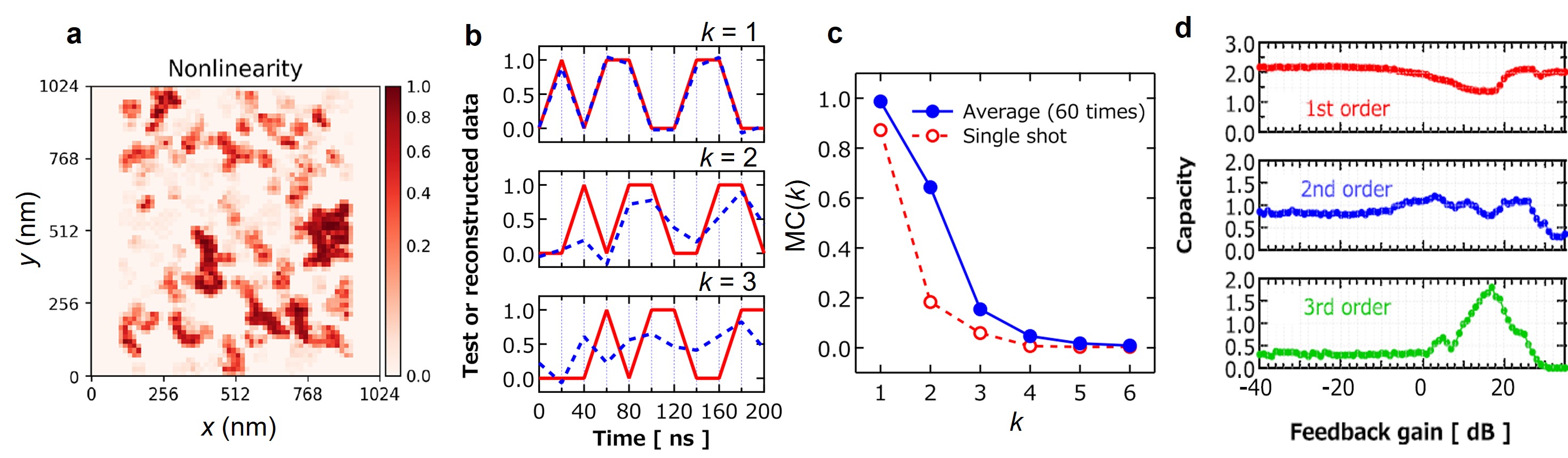}
  \caption{Reservoir computing metrics evaluations. \textbf{a}, Spatial analysis of NL for a computationally defined skyrmion reservoir with an electrical input. NL is distributed inhomogeneously due to the spatial distribution of magnetic skyrmions and their nonlinear dynamics excited by the current. \textbf{b},Comparisons between the test (red solid line) and reconstructed (blue dotted line) data with different $k$ values. \textbf{c}, MC($k$) obtained by a single shot (red dotted) and averaged data (blue solid) as a function of $k$ in physical reservoirs based on a STNO.  \textbf{d}, First-(top), second-(middle) and their-order(bottom) IPC of a physical reservoir based on spintronic oscillators, externally controlled by feedback gain. Panel \textbf{a} adapted with permission from Ref.~\citeonline{love2023spatial}, American Physical Society; Panels \textbf{b-c} adapted with permission from Ref.~\citeonline{Tsunegi_JJAP2018}, IOP Publishing; Panel \textbf{d} adapted with permission from ref.~\citeonline{Tsunegi_AdvIntSys2023}, Wiley-VCH GmbH. }
  \label{fig:reservoir1}
\end{figure}

\textbf{\textit{Non-Linearity (NL)}} assesses the reservoir's capacity to perform high-order non-linear transformations from input data, a critical feature for solving linearly inseparable problems\cite{Dambre_SciRep2012} and important for tasks such as data classification (e.g. handwritten digit recognition and image labelling) and temporal signal processing\cite{Du_NComm2017,Moon_NElec2019,Zhong_NComm2021,Torrejon_Nature2017,TANAKA_NeurNet2019}. While software-based reservoirs can be easily implemented with nonlinear (e.g. sigmoid) functions as a source of nonlinearlity, physical reservoirs require nonlinear responses to external stimuli.
To assess the nonlinearity, one approach is the method described by Love \textit{et al.}\cite{love2023spatial} in which a linear estimator $\hat{y}(t)$ is trained on an input time-series $u(t)$ as $\hat{y}(t) = \sum_{i} w_{i} u(t)$; here, the weights with the $i$th dimension $w_{i}$ are optimised during training. The optimised $w_{i}$ is used to generate $\hat{y}(t)$ for unseen input data $u(t)$, to compare with the real target value ${y}(t)$ for taking covariance ($\text{cov}(\hat{y}(t),y(t))=(1/n)\sum_{t}(\hat{y}(t)-\overline{\hat{y}(t)})(y(t)-\overline{y(t)})$ with the overbar representing the mean value and $n$ being the number of data points) and their respective variance ($\sigma^2$). The nonlinearity is calculated by:
\begin{equation}\label{eqn:nl}
\text{NL} = 1 - \frac{\text{cov}^{2}(\hat{y}(t),y(t))}{\sigma^{2}(\hat{y}(t))\sigma^{2}(y(t))}.
\end{equation}
The correlation term in the equation becomes zero(one) for a completely nonlinear(linear) output system, leading to $\text{NL}=1(0)$. This metric was exploited for assessing several physical reservoirs, including spatially resolved dynamics of simulated magnetic skyrmions~\cite{love2023spatial}(Fig.~\ref{fig:reservoir1}a) where NL was computed across different spatial regions and associated with the spatial distribution of magnetic skyrmions excited by electric currents. For readout, the spatially resolved out-of-plane magnetisation was used to perform computational tasks and extract reservoir metrics. The correlation between locally evaluated NL and distribution of magnetic skyrmions is discussed. NL analysis has also been applied to experimentally defined skyrmion reservoirs~\cite{Lee_NMater2024} and artificial spin ice\cite{Gartside_NNano2022,stenning2024neuromorphic} where the physical reservior state was modified to control metrics  including NL.   \newline

\textbf{\textit{Memory capacity}} (MC)~\cite{jaeger2002tutorial,love2023spatial} quantifies the reservoir's ability to retain information from past inputs, which is crucial for tasks involving temporal dependencies. MC is typically calculated by summing correlations between current reservoir states and lagged inputs:
\begin{equation}
\text{MC}(k) = \sum_{n} \frac{\text{cov}^{2}({u}(t-k),y(t))}{\sigma^{2}({u}(t-k))\sigma^{2}(y(t))},
\label{eq:MC}
\end{equation}
where the correlation between the input at time step \( t-k \) and the reservoir state at time step \(t\), which is summed up for the number of data points $n$ to obtain MC for each delay step $k$. $\text{MC}(k)$ can be further summed up for estimating the overall memory capacity. Decorrelation of input signals is essential to prevent overestimation of MC. To evaluate MC of reservoirs, one can input data generated by a random input function into the reservoir and calculate MC using Eq.~\ref{eq:MC}. Tsunegi \textit{et al.}~\cite{Tsunegi_JJAP2018} evaluated MC of a vortex-type STNO and demonstrated the correlation between the reservoir state and input data. They used the delayed time series data for $k$ = 1,2,3 to reconstruct the time series without delay ($k$ = 0) to assess the reservoir's memory property as shown in Fig.~\ref{fig:reservoir1}b. For $k$ = 1, the 20 ns delayed series (blue dotted curve) can excellently produce the targeted data (red solid line) whereas as the delay is increased, the reproduction quality of targeted data becomes increasingly worse. This can be quantified by MC($k$) that is plotted as a function of $k$ in Fig.~\ref{fig:reservoir1}c for both single-short and 60-times-averaged cases. A similar methodology of quantifying MC has been performed for various spin-based and non-spin-based reservoirs~\cite{Nakane_PRApp2021,Gartside_NNano2022,Lee_NMater2024,stenning2024neuromorphic}. While MC in general refers to the linear component of the fading memory property of reservoirs~\cite{jaeger2002tutorial}, the non-linear (higher-order) components can also be assessed using \textbf{\textit{informational processing capacity(IPC)}}~\cite{Dambre_SciRep2012,Duport_OptEx2012,Pauwels_FroPhys2019,Akashi_PRRes2020,Tsunegi_AdvIntSys2023} which characterises both linear and nonlinear reservoir moemries simultaneously. Figure~\ref{fig:reservoir1}d displays an example where high-order (nonlinear) memory components are strongly present and behave differently from the linear component as a function of feedback gain (reservoir hyperparameter)\cite{Tsunegi_AdvIntSys2023}. The correlation between MC and informational processing capacity and task-specific performance has been discussed e.g. in magnetic skyrmion reservoir~\cite{Lee_NMater2024} and delay coupled Stuart–Landau oscillators~\cite{Hülser_NanoPhton2023}. 

\textbf{\textit{Complexity\rev{/Computational Capacity}}} refers to a reservoir’s ability to generate diverse, linearly separable output states. To quantify this, metrics that measure the size of the latent space spanned by the outputs are needed. For example, applying a discrete random temporal signal to the reservoir and collecting the outputs into a square matrix allows one to define complexity as the effective rank of the matrix~\cite{roy2007effective}. Dimensionality reduction techniques such as principal component analysis are promising~\cite{gallicchio2010markovian,bianchi2020reservoir} to analyse the reservoir properties by identifying important reservoir states during dimensional reduction. 

\subsubsection*{Benchmarking tasks}

Physical reservoir performance has been quantified by the normalized mean-square error of benchmarking tasks. Since reservoir computing is primarily designed for processing time-series data due to its rapid learning advantages, computational tasks, such as nonlinear autoregressive
moving average with $n$th-order time lag (NARMA$n$)~\cite{Nakajima_SciRep2015,Duport_SciRep2016,Kanao_PRAppl2019,Akashi_PRRes2020,Kan_PRAppl2021,Namiki_AdvIntSys2023,Namiki_NeurCompEng2024,Iihama_npjspin2024} and chaotic Santa Fe or Mackey-Glass time series~\cite{Appeltant_SciRep2014,Sugano_IEEE2020,Gartside_NNano2022,Sun_ncom2023,Lee_NMater2024,Li_PRAppl2024,Zhou_SciAdv2025}, have been widely used for assessing reservoirs. NARMA2 requires a modest memory property has been used in several studies, both spin-based~\cite{Kanao_PRAppl2019,Akashi_PRRes2020,Namiki_NeurCompEng2024} and non-spin-based~\cite{Nakajima_SciRep2015,Kan_PRAppl2021,Namiki_AdvIntSys2023,Namiki_NeurCompEng2024}, whereas NARMA10 demands stronger memory performance from 10 steps apart, evaluated in studies using physical reservoirs~\cite{Paquot_SciRep2012,Duport_SciRep2016,Iihama_npjspin2024,Li_PRAppl2024}. Physical reservoir's computational capability has been tested by other machine learning tasks including signal transformation~\cite{Gartside_NNano2022,Lee_NMater2024} as well as classification tasks for recognising handwritten-digit~\cite{Yokouchi_SciAdv2022,Lee_SciRep2023,Namiki_AdvIntSys2023} and spoken-digit~\cite{Paquot_SciRep2012,Duport_OptEx2012,watt_arXiv2020, Msiska_AdvIntSys2023}. 
The correlation between benchmarking task results and task-independent metrics can be visualised by plotting the Spearman's rank correlation coefficient~\cite{spearmanBook} for physical reservoirs~\cite{Lee_NMater2024}.  
It is worth noting that performance tends to be significantly better for simulated reservoirs than that of experimental systems, due to a lack of experimental noise and drift in simulated systems; perfectly reproducible spatially-resolved dynamics can be readily simulated, which is often challenging experimentally. The reproducibility is a critical factor for RC due to its deterministic neural network architecture.

\subsubsection*{Challenges in Task-Independent Metric Characterisation}
It is essential to acknowledge that the metrics used to evaluate the computational performance or capacity of reservoirs can be influenced by the broader computing architecture, including the choice of input data and the mode of operation. When time delay is used to enhance non-linearity and complexity, an inherently linear system may exhibit a nonlinearity deviating from zero due to discretisation effects and instabilities induced by the time delay \cite{WANG_JDE2024}.
A linear system with time delay probed by white noise can potentially yield high nonlinearlity as the uncorrelated noise causes the synchronously measured covariance between predicted and actual (delayed) output to vanish. 
When assessing the distribution of memory across different delay steps and their linear/nonlinear components, waterfall plots can visualise the memory of each $t-k$ time-step delay for reservoir output-signal channels/nodes\cite{stenning2024neuromorphic}.
These underscore the importance of thoughtfully choosing input signals and operation mode, as inappropriate choices - much like in any machine learning algorithm - can lead to misleading performance evaluations.

\subsection{Magnetic Ising Machines}

\begin{figure}[ht]
    \centering
    \includegraphics[width=0.95\linewidth]{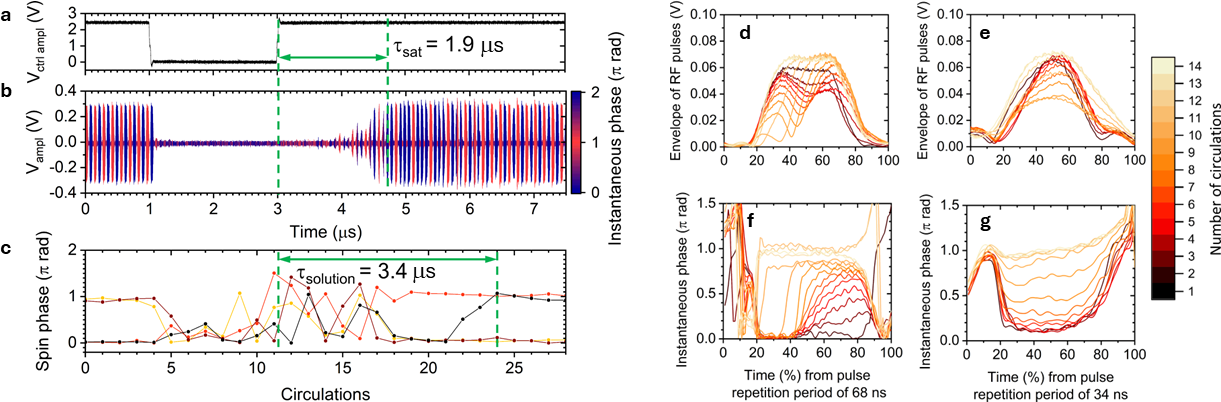}
  \caption{SWIM time traces for the measurement of time-to-saturation $\tau_{sat}$ and time-to-solution $\tau_{solution}$ parameters. \textbf{a} A control signal 
  turning on/off the SWIM. 
  \textbf{b} The propagating RF pulses 
  colored according to their instantaneous phase.
  \textbf{c} Phase values in the center of each propagating RF pulse sampled at each circulation period. 
  Spin switching scenarios for 4- and 8-spin MAX-CUT problem computations.
Envelopes (\textbf{d, e}) and instantaneous phase signals (\textbf{f, g}) of the 3rd propagating RF pulse signal within 14 circulation periods plotted in the form of overlapping traces in a relative scale for 4-spin (\textbf{d, f}) and 8-spin (\textbf{e, g}) MAX-CUT optimization problem. The color of the time traces corresponds to the circulation number. All figures reproduced from Ref.~\citeonline{litvinenko2023spinwave} with permission.}
  \label{fig:SWIM_1}
\end{figure}

Metrics directly related to solving COPs include the total number of spins, whether they exhibit all-to-all couplings or a more limited connectivity, the resolution of the coupling coefficients, and the time-to-solution. Metrics related to the IM hardware include energy efficiency, system size/footprint, temperature of operation, the associated need for 
cryogenic chambers or thermostats, and long-term stability sometimes requiring phase-locked loops for reference clocks and signals. 
Auxiliary metrics may include the expansion to higher-order couplings allowing one to solve larger problems  with a smaller number of Ising spins~\cite{bybee2023efficient}, the number of spin states going beyond binary representation, effectively converting an Ising machine into a Pott's~\cite{inoue2023coherent,inaba2022potts,bashar2024fpga} or hyper-spin machine~\cite{calvanese2024hyperscaling}, which, again, allows for the remapping of large Ising problems onto smaller size multi-state systems.  

\subsubsection*{Task-Independent Metrics}

\textbf{\textit{Amplitude relaxation and spin update time}} Oscillator-based Ising machines relax to an optimum solution via a gradual change of the oscillation amplitude or the phase. Spintronic oscillators, such as STNOs and SHNOs, are strongly nonlinear, non-isochronous microwave oscillators with a strong coupling between their oscillation amplitude and frequency. 
The rate of change of the oscillator states, and hence the final time-to-solution, then scales with 
the amplitude relaxation frequency parameter~\cite{Jiang_APRev2024}, which in SHNOs and STNOs operating at GHz frequencies can range from 100 MHz to 500 MHz, 
allowing them to change their amplitude and phase on the time scale of 2 ns to 10 ns. This time then corresponds to the minimum time required to obtain a single solution, 
$T_{single run}$.

In spin-wave time-multiplexed IMs, $T_{single run}$ 
scales with the delay of the spin-wave delay line, which defines a round trip time and an update rate for the equivalent Ising spins. $T_{single run}$ is obtained by multiplying the delay time with the number of relaxation periods, $N_{relaxation}$, or round trips required for the relaxation process towards a solution. Fortunately, $N_{relaxation}$ depends logarithmically on the size of the system with about 10-100 relaxation periods needed for a system of 100-1000 equivalent Ising spins. Taking into account the magnon lifetime, realistic delay times for spin-wave delay lines can be from hundreds of ns to single $\mu s$. Hence, the single solution time $T_{single run}$ in spin-wave IMs can be estimated to $1-10^3~\mu s$.

\textbf{\textit{Energy Consumption ($E$)}} takes into account the energy used during computation, including power consumed by the magnetic devices and supporting electronics over the computation time. SHNO-based Ising machines have potentially better energy consumption perspectives as the coupling between spins is passive. In contrast, spin-wave Ising machines require computation of matrix-vector multiplication for each round trip, 
usually done using an FPGA
~\cite{Honjo2021SciAdv100kCIM, cen2022large, bohm2019poor, litvinenko2024SAWIM}. Thanks to recent advances in spintronics, FPGA matrix multiplication blocks may be substituted with memristor-based matrix multiplication units ~\cite{tchendjou2022spintronic,li2018analogue} reducing the power consumption further. 

\textbf{\textit{Scalability ($N$)}} defines the maximum number of spintronic oscillators or propagating spin-wave RF pulses that can be inter-coupled in the system and serve as equivalent Ising spins. SHNO-based IMs are limited in terms of interconnection which can be traded off by remapping the Ising problem onto a King's graph in planar 2D SHNO-arrays with 4-coupling per oscillator. \rev{Minor-embedding algorithms map every logical spin to a ferromagnetically bound chain of physical oscillators, so a fully connected N-spin problem requires $N_{phys} = (N_{logical} - 1)^2$ physical oscillators~\cite{Hung2024Embedding,okuyama2016contractive}. Moreover, recent heuristics such as probabilistic-swap-shift annealing (PSSA) allow to reduce the overhead for sparse cubic or Barabási–Albert graphs~\cite{sugie2021minor}. We note that this} will significantly increase the number of physical oscillators required to emulate an equivalent \rev{logic} Ising spin. However, taking into account the very high consistency of SHNO parameters such as frequency current dependence, their robust synchronization properties~\cite{kumar2023robust} and, finally, nanoscale size, embedding large Ising problems onto extremely large SHNO arrays is a solvable engineering problem. For SWIMs, the dense all-to-all connectivity through digital coupling control using FPGAs make the system scalable to large spin states. With 
arrays of fabricated microscale YIG waveguides with lateral dimension of 2.66 mm×1.5 mm and each with its own linear and phase-sensitive amplifiers, SWIMs can be scaled to >100,000 spins~\cite{litvinenko2023spinwave}. 

\subsubsection*{Task-dependent Metrics}
\begin{figure}[ht]
    \centering
    \includegraphics[width=1.0\linewidth]{MagnIMbench.png}
  \caption{\textbf{Magnetic Ising machine benchmarking.} The table considers two distinct IM architectures -- spacial and time-multiplexed, parameters taken from corresponding references~\cite{dutta_experimental_2019,dutta2021ising,houshang2022phase,albertsson2021ultrafast,litvinenko2023spinwave,Honjo2021SciAdv100kCIM}. For comparison, we also include the noisy mean-field GPU algorithm, phase-transition oscillator IMs, and Coherent Ising Machines (CIMs).}
  \label{fig:MagIsingtable}
\end{figure}

\textbf{\textit{Probability of achieving a solution of a certain quality ($P_{\text{success}}$)}}; The probability $P_{\text{success}}$ of achieving a solution of a certain quality is unique to each particular problem and strongly depends on the Hamiltonian energy landscape, i.e. the number of local minima and their shape. However, statistically, $P_{\text{success}}$ is influenced by the density of the solved Ising matrix and the number of equivalent spins. $P_{\text{success}}$ is defined experimentally for each particular implementation of the Ising machine.

\textbf{\textit{Time-to-Solution at a Certain Quality Level ($TTS_{\chi}$)}}
Classical Ising machines often get trapped in a local suboptimal solution, especially when the system involves more than 10-100 equivalent Ising spins. Despite this, there are numerous combinatorial optimization tasks in which these suboptimal solutions are still valuable. Therefore, it is important to estimate the expected time required to obtain a solution for a random combinatorial problem that is within a certain percentage ($\chi$) of the optimal solution. This metric accounts for both the time per individual run ($T_{\text{single run}}$) and the probability ($P_{\text{success}}$) of achieving a solution of the desired quality in a single run. It is given by:

\begin{equation} 
TTS_{\chi} = T_{single run} \times \frac{\ln(1 - \chi/100)}{\ln(1 - P_{success})}. \label{eq}
\end{equation}

This metric takes into account the stochastic nature of classical IMs and allows estimation of the expectancy time for obtaining a suboptimal solution of certain quality $\chi$.

We benchmark the magnetic IM according to the previously described metrics (see Fig.~\ref{fig:MagIsingtable}) against similar technologies, including phase-transition oscillators ~\cite{dutta_experimental_2019, dutta2021ising} as an example of physical oscillator array IM, 100,000-spin optical coherent IM ~\cite{Honjo2021SciAdv100kCIM} as a state-of-the-art time-multiplexed IM and noisy mean-field GPU method ~\cite{king2018emulating} as a reference method exploiting von Neumann architecture computers. Note that most of the benchmarking parameters in the table are taken directly from the corresponding references ~\cite{dutta_experimental_2019, dutta2021ising,Honjo2021SciAdv100kCIM,houshang2022phase, albertsson2021ultrafast, king2018emulating} without additional calculations. Energy-to-solution parameter is a product of time-to-solution parameter and power consumption and has a dimension of Watts per solution. Energy efficiency is an inverse of energy-to-solution parameter and is measured in solutions per second per Watt. Magnetic IMs based on physical oscillator arrays demonstrate GHz oscillation frequencies while having continuous interaction, which potentially leads to the fastest single run times and overall time-to-solution parameters. For example, recent implementations with SHNOs and STNOs have oscillation frequency of 7.8 GHz~\cite{houshang2022phase} and 3.2 GHz ~\cite{albertsson2021ultrafast}. Typically, the amplitude relaxation frequency, i.e. a characteristic reaction frequency of the oscillator, is around an order of magnitude lower than the central frequency. Hence, spintronic oscillators can change their phase on a 10-100 ns time scale with a time-to-solution metric around a $\mu$s. However, because of 2.5D nanofabrication technology and the difficulty of implementing an all-to-all interconnection in an array of oscillators nanofabricated in a single layer, the scalability of SHNO and STNO-based IMs is still an issue. Time-multiplexed spin-wave Ising machines provide a clear pathway for large-scale IMs but offer a relatively slow time-to-solution parameter in the range of tens to hundreds $\mu s$ due to the low interaction frequency of 3.69 MHz between the spin-wave propagating oscillators. The interaction frequency is an inverse of the loop delay time and describes how frequently the pulses are updated with an FPGA-based measurement and feedback block. SHNO- and STNO-based IMs offer the best power consumption similar to phase transition oscillators. The fundamental reason is their low nonlinear thresholds, which enable nanoscale oscillators to reach nonlinear saturation at low signal amplitudes, resulting in power consumption in the range of a few mWatts. In contrast, nonlinear phase-sensitive amplification in SWIM and coherent IM consumes power on the order of watts~\cite{Honjo2021SciAdv100kCIM}.

\section{Perspective/conclusions}

\rev{Building on the rapid progress of spin-based computing as discussed in this Technical Review, fundamental technological gaps remain between the current state of the art and its successful realisations. Across the overarching spin-based approaches, there exist widely opened areas of research for the development of co-integration with CMOS  (CMOS+X). Here, key challenges include identification of the competitive functionalities of spin-based devices against mature CMOS counterparts, interconnect scaling, energy-efficient coupling mechanisms and their successful monolithic integration fabrication processes. Hardware-software co-design will potentially help mitigate device-to-device variability of spintronic devices. In what follows, we offer our future perspective view.}

Neuromorphic computing exploits a physical system as a part of the entire architecture, with the rest supported by additional digital circuitry and software to complete computation. How much the digital/software components can be effectively replaced with  hardware is an ongoing challenge. For example, the majority of RC demonstrations performed a regression process on external computer software which should be soon replaced by hardware processing units to realise fully connected, on-chip recurrent neural networks. To this end, an efficient use of magnetic and spintronic interactions (such as spin-wave interaction and spin transfer torques) for data-transfer and processing can be actively studied, \rev{removing unscalable magnetic field protocols as well as . Further research on efficient on-chip time-multiplexing for high-dimensional mapping in physical reservoirs must be carried out to develop efficient help mitigate further circuitry complexity for such data processing.}   

\rev{While the electrical coupling between spin-based and CMOS devices has been already demonstrated with promising results in probabilistic Ising machines~\cite{borders2019integer, si2024energy} and energy harvesting~\cite{Sharma_nelec2024},  it is important to recognise that energy costs from peripheral circuitry (e.g., amplifiers in RF neurons or multiply and accumulate operation in p-bit systems) can potentially become the dominant source of the energy consumption~\cite{sutton2020autonomous}.} Metrics, such as energy per synaptic operation and power consumption per neuron, and specific benchmarks will allow direct comparisons of spin-based approaches with other technologies, such as photonic or resistive computing while taking into account the role of the peripheral circuitries.

Hence, a key advance and challenge will be to identify and develop standardized manufacturing protocols for co-integrating spin-based devices with existing CMOS technology. This development will be accelerated by simulation tools to design hybrid spintronic-CMOS systems where co-designing algorithms are tailored to the physical properties of spintronic devices. For memory applications, MTJs have successfully entered into the back-end-of-line of CMOS processes but their full potential has been limited by fabrication challenges, thermal stability, and large device-to-device variability. We expect that spin-based neuromorphic technologies at high technology readiness levels (TRLs) will also face similar technological challenges, for which material innovation and device design will remain a central approach. For example, sophisticated MTJ stacks, such as those with a synthetic antiferromagnetic structure for both free and fixed layers might offer better p-bit functionality. Spin-based computing would benefit from hybrid designs using other order parameters such as that in ferroelectricity which has already shown a promising electric-field control of MTJ states~\cite{Chen_ncomm2019,Sun_advsci2024}. Antiferromagnet materials are another exciting material family with spin-based THz technology~\cite{Wu_JAP2021}. 


\rev{Magnetic p-bits provide a scalable approach to probabilistic computing, with potential as efficient platforms for generative AI, machine learning, sampling, optimization, inference, and quantum simulation. To realize these goals, several challenges remain for viable large-scale implementations, including increasing the speed of random number generation, reducing the power to generate random numbers, and improving stability against magnetic field perturbations and temperature variations, as described in Sec.~3.2. Importantly, these challenges must be addressed not only through device engineering but also at the circuit, architectural, and algorithmic levels. Furthermore, the development of synaptic circuits, countermeasures for bit-to-bit variability, and monolithic integration will be essential for scaling up to million-p-bit systems. Nonetheless, the performance demonstrated by digital implementations at the 10,000 to 50,000 p-bit scales~\cite{aadit2022massively, nikhar2024alltoall, niazi2024cmos} highlights the promise of larger systems that fully harness the potential of the s-MTJ concept.}

Beyond s-MTJs, several alternative physical systems have been explored for implementing probabilistic bits. Voltage-controlled MTJs provide a pathway for energy-efficient probabilistic switching by leveraging anisotropy modulation through applied voltages~\cite{WOS:000912390600001}. Another promising avenue involves STNOs, where phase noise is harnessed as a source of stochasticity for probabilistic computing~\cite{PhysRevApplied.21.034063}. Additionally, non-magnetic solid-state implementations such as resistive RAM and memristors have demonstrated intrinsic stochastic switching behavior, making them viable candidates for probabilistic computing architectures~\cite{woo2022probabilistic}. Other CMOS-compatible approaches, including stochastic elements based on avalanche and Zener diodes, have also been proposed as efficient sources of random number generation~\cite{whitehead2023cmos, patel2024pass}.

IMs based on GHz frequency spintronic oscillators provide the fastest time-to-solution parameter amongst other architectures due to their continuous inter-oscillator interaction and extremely fast amplitude relaxation rates. The main challenge for these computational systems is the scalability in terms of number of equivalent spins. SHNOs are principally planar devices that demonstrate controllable and effective inter-oscillator coupling via dipolar fields ~\cite{erokhin2014robust,chen2016phase} and propagating spinwaves ~\cite{kendziorczyk2014spin,houshang2016spin,kendziorczyk2016mutual,kumar2025spin}. In planar topology, the coupling is limited to the nearest neighbors. STNOs exhibit relatively high output power and can be effectively coupled via electrical connections ~\cite{locatelli2015efficient,sharma2021electrically} but due to the GHz frequency range and significant cross-talk parasitic coupling the number of these connection and their density has to be limited as well. Nevertheless, complex graphs can still be embedded onto SHNO/STNO-based IMs built with the King's graph structure ~\cite{lo2023ising} with a certain overhead in terms of the ratio between the number of physical oscillators and the number of equivalent logical spins. Another challenge that spintronic oscillators face is the speed of the read-out of the solution that is limited by the low signal-to-noise ratio. Hence, orders of magnitude improvement in terms of output power is crucial for the speed performance of the future SHNO/STNO-based IMs.

The use of propagating spin-waves with extremely slow group velocities allows the miniaturization of time-multiplexed IMs, potentially opening up a way for on-chip integration. However, in order to inherit the scalability potential of thousands of equivalent Ising spins typical for time-multiplexed systems, it is necessary to cope with the problem of nonlinear dispersion that limits the minimal temporal width of equivalent spins in SWIM's delay lines. Fortunately, there are physical and engineering techniques that allow one to achieve nearly dispersion-less propagation of spin-waves including the exploitation of dipole-exchange hybridization ~\cite{divinskiy2021dispersionless} and combination of different magnetization schemes ~\cite{fetisov1998disp,sethares1980msw}. Moreover, there is a potential to substitute an electrical phase sensitive amplifier with spin-wave amplification techniques ~\cite{merbouche2024true} eliminating the need for highly inefficient electromagnetic-to-spin-wave transducers and opening up the pathway for all-spintronic SWIM implementation and even more miniature design.

In conclusion, it is expected that spin-based paradigms will offer unique advantages in energy efficiency, scalability, and task-specific optimization and considering their rich nonlinear dynamics, stochasticity, time-non-locality and the already demonstrated compatibility with CMOS processes. A future direction lies in combining these paradigms in hybrid approaches. For instance, the integration of reservoir computing with probabilistic inference could enhance the capabilities of generative models, also the combination of different implementation of magnetic Ising machines can improve and speed-up the search of solution of constrained optimization problems. \rev{To correctly evaluate their computing capacity/performance, metrics introduced and discussed in this Technical Review should be effectively used for benchmarking each physical system as signposts.} Spintronic systems are not just an alternative but a necessity for the future of computing.

\bibliographystyle{naturemag-doi}
\bibliography{nrevphys_neurospin}

\subsection*{Acknowledgements} 
H.K. thanks the Leverhulme Trust for financial support via their Research Fellowship (RF-2024-317) \rev{and JSPS for their support through Kakenhi (Grant no. 25H00837)}. H.K. and J.C.G. were supported by EPSRC grant EP/X015661/1. J.C.G. was supported by EPSRC grant EP/Y003276/1 and the Royal Academy of Engineering Fellowship RF2122-21-363.
A.L. acknowledges funding from the Marie Skłodowska-Curie grant agreement No.~101111429 "SWIM". J.Å. acknowledges funding from the Horizon 2020 research and innovation program No. 835068 "TOPSPIN" and as a Swedish Research Council Distinguished Professor (Dnr.~2024-01943). 
K.E.-S. acknowledges funding from the Emergent AI Center funded by the Carl Zeiss Foundation and the German Research Foundation (DFG) Project‐ID 403233384 (SPP Skyrmionics) and 405553726 (CRC/TRR 270, project B12). S.F. acknowledges funding from JST-ASPIRE (Grant no. 
JPMJAP2322), JST-CREST (Grant no. JPMJCR19K3), and JSPS Kakenhi (Grant nos. 24H00039, 24H02235, and 25H00447). K.Y.C and K.S. acknowledge support from the Office of Naval Research (ONR), Multidisciplinary University Research Initiative (MURI) grant N000142312708.
\rev{T.T. acknowledges JSPS Kakenhi (Grant no. 24K01336). }
The authors thank Julie Grollier and Frank Mizrahi for their help on the initial stage of manuscript writing and Mark Stiles and Daniel Gopman for their helpful feedback. The authors acknowledge Maria Azhar for drawing the skyrmion image in Fig.~1c.  

\subsection*{Author contributions}
H.K., G.F. and S.F. initiated this project and all authors contributed to the preparation of the article.

\subsection*{Competing interests}
The authors declare no competing interests.

\newpage
\section*{Supplementary Table 1: Metrics of magnetic reservoir systems}

\begin{table}[h]
    \centering
    \scriptsize  
    \renewcommand{\arraystretch}{0.9} 
    \setlength{\tabcolsep}{2pt} 
    \begin{tabular}{p{0.5cm} p{2.8cm} p{1.7cm} p{0.9cm} p{0.9cm} p{0.9cm} c c p{2.5cm} p{2.5cm} p{2.5cm}}
        \toprule
        \textbf{Ref} & \textbf{Physical System} & \textbf{Sim. or Exp.} & \textbf{\# Phys. Nodes} & \textbf{\# Virt. Nodes} & \textbf{\# Total States} & \textbf{MC} & \textbf{NL} & \textbf{Task} & \textbf{Performance} & \textbf{Notes} \\
        \midrule
        \citeonline{Torrejon_Nature2017} & Spintronic Oscillator & Experiment & 1 & 400 & 400 & - & - & Spoken digit recognition & Accuracy = 99.6\% & - \\
        \citeonline{Tsunegi_AdvIntSys2023} & Spintronic Oscillator & Experiment & 1 & 200 & 200 & 2.1 & - & NARMA 2 & NMSE = 4$\times$10$^{-6}$ & Also assess nonlinear memory via IPC, IPC score of 5.6. NMSE definition A \\
        \citeonline{Kanao_PRAppl2019} & Spintronic Oscillator Array (Dipolar Coupled) & Simulation & 1600 & - & - & 56 & - & NARMA 2 & NMSE = 8$\times$10$^{-8}$ & NMSE definition A 
        \\
        \citeonline{Gartside_NNano2022} & Artificial Spin Ice array & Experiment & 400 & 1 & 400 & 5.8 & 0.4 & Mackey Glass t+10 prediction & MSE = 9.936$\times$10$^{-3}$ & - \\
        \citeonline{stenning2024neuromorphic} & Network of Artificial Spin Ice Arrays & Experiment & 40000 & 1 & 40000 & 6.8 & 0.75 & Mackey Glass t+10 prediction & MSE = 4$\times$10$^{-3}$ & - \\
        \citeonline{vidamour2023reconfigurable} & Magnetic Nanoring Array & Experiment & 1 & 32 & 32 & 12 & - & NARMA 5 & NMSE = 0.265 & NMSE definition B \\
        \citeonline{vidamour2022quantifying} & Magnetic Nanoring Array & Simulation & 1 & 150 & 150 & 2 & - & Spoken digit recognition & Accuracy = 97.7\% & - \\
        \citeonline{Lee_NMater2024} & Chiral/Skyrmion magnet Cu$_2$OSeO$_3$ & Experiment & 1 & 1000 & 1000 & 7 & 0.71 & Mackey Glass t+10 prediction & MSE = 3.7$\times$10$^{-3}$ & - \\
        \citeonline{taniguchi2022spintronic} & MTJ & Simulation & 1 & 250 & 250 & 3 & - & NARMA 2 & NMSE = 8.43$\times$10$^{-6}$ & NMSE definition A \\
        \citeonline{Zhou_SciAdv2025} & Magnetic thin film memristors - PtMn/CoFeB & Experiment & 14 & 6 & 84 & - & - & Spoken digit recognition & Accuracy = 91.6\% & - \\
        \citeonline{Iihama_npjspin2024} & Propagating spin waves in magnetic thin film & Simulation & 8 & 8 & 64 & 50 & - & NARMA 10 & NRMSE = 0.25 & Also assess nonlinear memory via IPC, IPC score of 75 \\
        \citeonline{watt2021enhancing} & Propagating spin waves in magnetic thin film & Experimental & 1 & 50 & 50 & 4.68 & - & - & - & - \\
        \citeonline{watt2020reservoir} & Propagating spin waves in magnetic thin film & Experimental & 1 & 20 & 20 & 3 & - & - & - & - \\
        \citeonline{nakane2018reservoir} & Propagating spin waves in magnetic thin film & Simulation & 3 & 1 & 3 & - & - & Spin wave pulse duration estimation & RMSE = 0.3 & - \\
        \citeonline{Namiki_AdvIntSys2023} & Propagating spin waves in magnetic thin film & Experimental & - & 196 & - & - & - & NARMA 10, NARMA 2 & NMSE = 0.168, 1.81$\times$10$^{-2}$ & NMSE definition B \\
        \bottomrule
    \end{tabular}
    \caption{Comparison of different physical reservoir computing systems. Due to the many different ways different researchers benchmark and assess reservoir systems, direct comparison is challenging and this table is not intended to rank systems in term of their performance, but rather provide a collated comparison of various physical systems, node arrangement, and metric and task assessment approaches. N.B. different researchers use different definitions of the normalised root mean square error (NMSE) but call it the same name. To aid comparison, we refer to definition A as in Ref \citeonline{taniguchi2022spintronic} where the error is normalised by dividing by the square of the target data, and definition B as in Ref \citeonline{vidamour2023reconfigurable} where the error is normalised by dividing by the square of the variance of the reservoir's reconstructed signal, e.g. its weighted computational output.}
    \label{tab:comparison}
\end{table}

\newpage
\section*{\rev{Supplementary Note 1: Further key metrics for assessing reservoir computing systems}}

\rev{We here list further key metrics that are often used in the wider reservoir computing community.}\\

\noindent \rev{\textbf{\textit{Kernel Rank }} captures how richly the reservoir maps inputs into linearly separable internal states, reflecting its capacity for memorization and pattern separation.}\\

\noindent \rev{\textbf{\textit{Generalization Rank }} indicates how consistently the reservoir responds to similar inputs, providing a measure of its robustness and generalization capability.}\\

\noindent \rev{\textbf{\textit{Lyapunov exponent}} To quantify the robustness of a reservoir, Lyapunov exponents are suitable. They indicate whether the system operates in a stable, chaotic, or edge-of-chaos regime. A positive Lyapunov exponent suggests chaotic behaviour and sensitivity to input noise, while a negative exponent indicates convergence to fixed points or attractors, limiting dynamic richness. Operating near zero (the edge of chaos) often provides an optimal balance between stability and computational richness.}\\

\noindent \rev{\textbf{\textit{Robustness}} refers to a physical implementation's ability to maintain stable and accurate performance despite variations in hardware fabrication, environmental noise, or device degradation over time. Key strategies to enhance robustness include implementing diverse reservoir designs, using tunable nonlinear dynamics and customizable timescales, and developing adaptive online learning methods to compensate for device imperfections. Techniques like linear field calibration and thermal noise reduction through frequency filtering also contribute to more reliable and practical physical reservoir computing systems.}\\

\noindent \rev{\textbf{\textit{Information Processing Capacity (IPC)}} provides a task-independent framework to evaluate the computational capability of a reservoir, including physical reservoirs. It measures how accurately a linear readout can reconstruct a set of predefined nonlinear functions of the input history (e.g., delayed inputs, polynomial terms). By summing the squared correlation between the predicted and target functions, IPC quantifies both memory and nonlinear transformation capacities. This approach is particularly useful in physical reservoir computing, where internal dynamics are often opaque, enabling standardized benchmarking of diverse physical substrates based solely on input-output behaviour.}\\

\end{document}